\documentclass[pdflatex,sn-mathphys-num]{sn-jnl}

\usepackage{graphicx}%
\usepackage{multirow}%
\usepackage{amsmath,amssymb,amsfonts}%
\usepackage{amsthm}%
\usepackage{mathrsfs}%
\usepackage[title]{appendix}%
\usepackage{xcolor}%
\usepackage{textcomp}%
\usepackage{manyfoot}%
\usepackage{booktabs}%
\usepackage{algorithm}%
\usepackage{algorithmicx}%
\usepackage{algpseudocode}%
\usepackage{listings}%
\usepackage{caption}%
\usepackage{subcaption}%

\DeclareMathOperator{\E}{\mathbb{E}}
\DeclareMathOperator{\pt}{\mathnormal p_{\mathrm T}}
\DeclareMathOperator{\Et}{\mathnormal E_{\mathrm T}}
\DeclareMathOperator{\Etem}{\mathnormal E_{\mathrm T, em}}
\DeclareMathOperator{\Ethad}{\mathnormal E_{\mathrm T, had}}
\DeclareMathOperator{\Etpred}{\mathnormal E_{\mathrm T, pred}}
\DeclareMathOperator{\ptone}{\mathnormal p_{\mathrm T, 1}}
\DeclareMathOperator{\ptfour}{\mathnormal p_{\mathrm T, 4}}
\DeclareMathOperator{\pu}{\langle\mu\rangle}

\newcommand{\Ht}{$H_{\mathrm T}$}
\newcommand{\MHt}{$H_{\mathrm T}^\mathrm{miss}$}

\newcommand{\ggfhh}{ggF $HH\rightarrow b\bar{b}b\bar{b}$}
\newcommand{\vbfhh}{VBF $HH\rightarrow b\bar{b}b\bar{b}$}
\newcommand{\ttbar}{$t\bar{t}\rightarrow b\bar{b}q\bar{q}q\bar{q}$}
\newcommand{\hzbbvv}{$HZ \rightarrow b\bar{b}\nu\bar{\nu}$}

\theoremstyle{thmstyleone}%
\theoremstyle{thmstyletwo}%

\theoremstyle{thmstylethree}%

\begin{document}

\title[Article Title]{End-to-end optimisation of HEP triggers}

\author*[1,2]{\fnm{Noah} \sur{Clarke Hall}}\email{noah.clarke-hall.22@ucl.ac.uk}
\equalcont{These authors contributed equally to this work.}

\author[2]{\fnm{Ioannis} \sur{Xiotidis}}\email{ioannis.xiotidis@cern.ch}
\equalcont{These authors contributed equally to this work.}

\author[1,2]{\fnm{Nikos} \sur{Konstantinidis}}\email{n.konstantinidis@ucl.ac.uk}

\author[2,3]{\fnm{David W.} \sur{Miller}}\email{david.w.miller@uchicago.edu}

\affil*[1]{\orgdiv{Department of Physics and Astronomy}, \orgname{University College London (UCL)}, \orgaddress{\street{Gower Street}, \city{London}, \postcode{WC1E 6BT}, \country{UK}}}

\affil[2]{\orgdiv{Experimental Physics}, \orgname{CERN}, \orgaddress{\street{Espl. des Particules 1}, \city{Meyrin}, \postcode{1217}, \country{Switzerland}}}

\affil[3]{\orgdiv{Department of Physics, Enrico Fermi Institute, Kavli Institute for Cosmological Physics}, \orgname{University of Chicago}, \orgaddress{\street{Ellis Ave.}, \city{Chicago}, \postcode{60637}, \state{Illinois}, \country{USA}}}


\abstract{
High-energy physics experiments face extreme data rates, requiring real-time trigger systems to reduce event throughput while preserving sensitivity to rare processes. 
Trigger systems are typically constructed as modular chains of sequentially optimised algorithms, including machine learning models.
Each algorithm is optimised for a specific local objective with no guarantee of overall optimality.
We instead formulate trigger design as a constrained end-to-end optimisation problem, treating all stages—including data encoding, denoising, clustering, and calibration—as components of a single differentiable system trained against a unified physics objective. 
The framework jointly optimises performance while incorporating physics and deployment constraints. 
We demonstrate this approach on a hardware multi-jet trigger inspired by the ATLAS High-Luminosity Large Hadron Collider design. 
Using Higgs boson pair production as a benchmark, we observe x2-4 improvement in true-positive rate at fixed false-positive rate, while preserving interpretable intermediate physics objects and monotonic calibration constraints. 
These results highlight end-to-end optimisation as a practical paradigm for next-generation real-time event selection systems.
}




\maketitle

\section{Introduction}\label{sec1}
Modern High-Energy Physics (HEP) experiments operate at extreme data rates that rival or exceed those of the largest commercial data-processing systems. At the LHC~\cite{Evans:2008zzb}, detector ``snapshots", known as events, are recorded at the bunch-crossing rate of 40 MHz. Each event contains detailed information from tens of millions of detector channels, resulting in raw data rates of hundreds of terabytes per second. Storing or even fully reconstructing this data stream is technologically and economically unfeasible.
To manage this bandwidth, experiments such as ATLAS~\cite{ATLAS:2023dns} and CMS~\cite{Safonov:2010qg} employ hierarchical real-time filtering systems known as triggers. These systems progressively reduce the event rate by several orders of magnitude--from 40 MHz to O(1 kHz) for permanent storage.
This reduction is achieved through a multi-level architecture. The first stage (Level-1 trigger) is implemented in custom hardware, typically FPGA-based boards, and operates with fixed latencies of order microseconds. At this stage, only partial or coarse detector information can be used, and algorithms must satisfy stringent constraints on resource usage and determinism. Events accepted at Level-1 are passed to a high-level trigger (HLT), implemented in large CPU/GPU computing farms, where more detailed reconstruction and refined selection criteria can be applied under millisecond-scale average latency budgets. This bandwidth hierarchy--hardware-level, ultra-low-latency filtering, followed by software-level, higher-complexity processing--is fundamental to the data-acquisition model of modern collider experiments. The challenge for trigger systems is to achieve the above event rate reduction while maximizing the selection efficiency for interesting and relatively rare processes in an inclusive and unbiased manner. This is a fundamental factor in determining the discovery potential of the experiments.

Traditionally, trigger systems have been constructed as sequential processing chains. Individual algorithms (e.g. for denoising, clustering, calibration, or object identification) are designed and optimised independently under local performance criteria. Interpretability and robustness are maintained by propagating well-defined intermediate physics objects between stages. Even as machine-learning (ML) methods increasingly replace handcrafted heuristics within individual components~\cite{Astrand:2025sij}, the global architecture typically remains modular and sequential.
In contrast, modern machine-learning systems in industry and data science are commonly optimised end-to-end: entire computational graphs are trained jointly with respect to a single objective function. The tension between these paradigms—sequential, modular optimisation under hard latency, bandwidth and resource constraints versus holistic, end-to-end learning—defines a central methodological challenge for next-generation real-time data processing at the LHC.

In this paper, we formulate trigger design as a constrained end-to-end optimisation problem, treating each algorithm as trainable components of a single chain.
This end-to-end framework maximises the physics performance while still ensuring that intermediate physics objects are physical and interpretable.
Inspired by modern deep-learning techniques, we implement the entire trigger within a differentiable programming framework. 
We summarise physics performance as a single loss term which can then be minimised using gradient descent on all trigger parameters simultaneously. 
We show 
that the optimal overall design need not correspond to optimal performance for any intermediate algorithm.
We extend the end-to-end approach to a framework to include the joint optimisation of multiple trigger observables, and incorporate auxiliary tasks such as event reconstruction and object calibration.
Hardware constraints are also included in the optimisation, showing that data-encoding and machine-learning model-compression rules can also be optimised under a physics objective.

We note that the term ``end-to-end" has been used to describe data-processing in HEP in the past \cite{Andrews:2018nwy} \cite{Andrews:2019wng} \cite{Metzger_2025} \cite{Bhattacharya_2023}, but this refers to large single neural network architectures that operate directly on low-level detector data. 
These studies seem to target offline event selection and reconstruction, and do not meet the robustness, interpretability and hardware requirements for real-time event-selection.
Likewise, simultaneous classification and calibration has been proposed within an LHC trigger context \cite{Carlson:2025sso}.
However, calibration and classification are optimised independently by separate algorithms, and the two outputs are combined in a way that does not respect proper scoring rules.
This therefore falls short of the end-to-end optimal framework presented here.

Techniques for meeting hardware constraints have been extensively explored in a HEP trigger context.
The majority of work has targeted model-compression \cite{Aarrestad_2021}\cite{Coelho_2021}\cite{Govorkova_2022}\cite{zipper2023testingneuralnetworkanomaly}\cite{CMS-DP-2023-086}\cite{Cohen:2938881}, with the optimisation of data-encoding rules relatively under-explored \cite{Guglielmo_2021}\cite{Chen:2641465}.
However, none of these works resolve these constraints within a end-to-end physics-optimal framework.

We apply the end-to-end framework to a realistic hadronic jet trigger inspired by the ATLAS hardware trigger for the High-Luminosity Large Hadron Collider (HL-LHC) \cite{tdr}.
The trigger targets multi-object (multi-jet) events, including Higgs boson pair production, which is considered one of the top priorities for study by the European Strategy for Particle Physics \cite{EuropeanStrategyGroup:2020pow}. 
Our optimisation framework yields substantial gains in signal efficiency relative to sequential optimisation, while preserving robustness and interpretability requirements.
Overall, these results suggest that trigger systems can be treated as unified, task-aware systems, redefining how real-time event selection is designed in HEP.

The structure of the paper is as follows; Section \ref{sec:theory} outlines the optimisation framework and design choices necessary for end-to-end optimisation.
Section \ref{sec:experiments} details the application of our framework to a hardware trigger at the HL-LHC.
Finally, Section \ref{sec:discuss} discusses how this work can be extended and applied more widely in HEP experiments and beyond.

\section{Optimising triggers}\label{sec:theory}
\subsection{Trigger optimisation}
\label{subsec:trig_opt}
Let an event be represented by some granular detector data $x$ and hidden event label $c$.
Let $(x, c)$ be samples from some event-generating process $p(x, c)$.
We summarise a generic trigger as a one-dimensional discriminant $y=y(x; \theta)$, where $\theta$ is a set of parameters.
The Bayes-optimal trigger $\hat{y}=y(x;\hat{\theta})$ is the minimiser of the expected risk $\mathcal{R}$ \cite{vapnik1991risk}
\begin{equation}
    \hat{\theta} = \arg\min_\theta [\mathcal{R}],
\end{equation}
where
\begin{equation}
    \mathcal{R} = \E_{x, c \sim p(x, c)}[\mathcal{L}(y(x))]
\end{equation}
and the loss function $\mathcal{L}$ summarises performance for a specified task. 
We henceforth refer to the Bayes-optimal trigger as ``task-optimal".
Sampling batches of events as $x_i, c_i \sim p(x, c)$, we can calculate the empirical risk
\begin{equation}
\hat{\mathcal{R}} = \frac{1}{N} \sum_{i=1}^N \mathcal{L}\left(y(x_i)\right), 
\end{equation}
where $N$ is the batch size, and $\hat{\mathcal{R}}$ is an unbiased 
estimator of $\mathcal{R}$.
Automatic differentiation \cite{JMLR:v18:17-468} in modern machine-learning libraries allow the gradients $\partial \hat {\mathcal{R}}/\partial\theta$ to be calculated simultaneously for all $\theta$.
While $y$ is piece-wise differentiable for all $\theta$, repeated batch sampling and stochastic gradient descent \cite{LeCun2012} on $\theta$ converges to the task-optimal estimator $\hat y$ under standard assumptions \cite{garrigos2024handbookconvergencetheoremsstochastic}.
\begin{figure}[h!]
    \centering
    \includegraphics[width=0.9\linewidth]{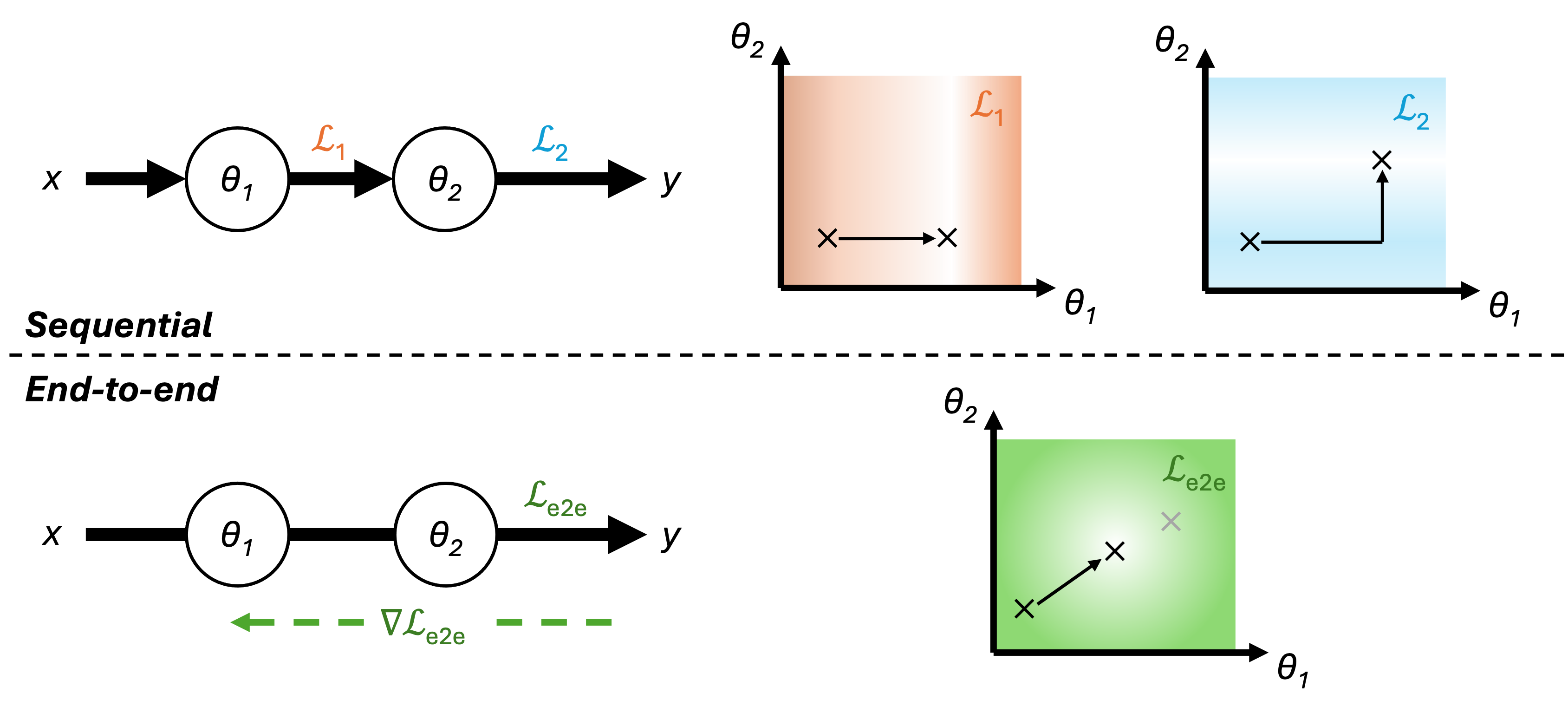}
    \caption{Comparison of optimisation approaches - sequential optimisation on $\mathcal{L}_1$ then $ \mathcal{L}_2$ generally fails to select the $\mathcal{L}_{e2e}$-optimal solution.}
    \label{fig:seq_vs_e2e}
\end{figure}

In this paper we compare two approaches to optimisation, sequential and end-to-end.
In the sequential case, each algorithm is optimised under an algorithm-specific objective.
Each optimisation selects a solution that this is locally optimal but mis-specified under the overall physics objective. 
Mis-specification compounds with each optimisation. 
The end-to-end optimisation uses simultaneous descent on all parameters under the overall physics objective, converging on the globally optimal result.
Fig. \ref{fig:seq_vs_e2e} shows a simplified trigger with two algorithms, each with a single parameter $\theta_1, \theta_2$ and specific objective $\mathcal{L}_1, \mathcal{L}_2$.
Let $\mathcal{L}_{e2e}$ be the overall physics objective.
The diagram shows how sequential optimisation on $\mathcal{L}_1$ then $ \mathcal{L}_2$ will in general fail to converge on the $\mathcal{L}_{e2e}$-optimal solution.

\subsection{Joint classification \& reconstruction}
\label{subsec:calibration}
The physics objectives of a HEP trigger are event classification and reconstruction.
The output of most classifiers is a unitless score, often a measure of probability such as a logit or the posterior $p(y|x)$.
However, reconstruction requires that the measured value of $y$ be regressed to some identifiable truth value $t$. 
For example, $y$ could be the reconstructed momentum of a particle and $t$ could be the true momentum. 
We define a joint loss function containing a classification term $C$ and a reconstruction term $D$, where $D$ represents the distance between $y$ and $t$, such that
\begin{equation}
\label{eq:bce_w_calib}
    \mathcal{L} = (1-\alpha)C[f_\phi(y)] + \alpha D(t,y),
\end{equation}
where the positive coefficient $\alpha\in[0,1]$ parametrises the relative strength of the two terms. 
Setting $\alpha=0$ produces the optimal classifier, whereas $\alpha=1$ produces the optimal event-reconstructor.
We note that the $\alpha=1$ case could be viewed as solving the ``particle-flow" problem within a trigger context \cite{Petrucciani:2650974}.

We resolve the difference in scale between $y$-space and score-space by introducing a learnable monotone bijection $f_\phi$ with parameters $\phi$. 
The transformation $y\rightarrow y'=f(y)$ preserves discrimination power, while decoupling $C$ and $D$ during optimisation and allowing $y$ to remain physical. 
Assuming sufficient capacity of the estimators $y$ and $f$, this decoupling allows the conditional reconstruction bias $\E[D|t]\rightarrow0$ for any choice of $\alpha>0$, since any effect from changing $\alpha$ can be fully absorbed by the freedom in $f$.
However, capacity constraints in $y$ such as small model size, monotonicity and continuity can reintroduce coupling and create conditional bias.
The location and size of these biases depend on both the specific constraint and $\alpha$.
The conditional variance $\text{Var}[D|t]$ also depends strongly on $\alpha$.

\subsection{Inclusive triggers}
\label{subsec:incl_trig}
Triggers, such as the latency bound hardware triggers, often need to prioritise inclusivity for a range of signals, rather than exclusively targeting a single well-defined signal model. 
For this reason, many LHC triggers place thresholds on individual compact physics objects, such as single clusters $x_\text{cluster} \subset x$.
In this case, the task-optimal solution for each trigger is determined only by the marginal posterior $p(y|x_\text{cluster})$.
These single-object triggers are highly inclusive, since they ignore signal-specific topological features, and instead exploit cluster-level features which are shared by a diverse set of signals.
Topological triggers that exploit correlations between objects are more brittle under changes in signal model, especially within a machine-learning context.
We therefore focus on optimising single-object triggers. 

\subsection{Multi-trigger optimisation}
\label{subsec:multi_trig}
Most HEP experiments have broad physics programmes. 
The trigger system reflects this, placing thresholds on multiple triggers $\{y_j(x)\}$, each targeting a different event topology.
 
If $\{y_j\}$ are conditionally independent, then the joint posterior factorises $p(\{y_j\}|x) = \prod_j p(y_i|x)$ and we can find $\{\hat{y}_j\}$ by minimising the expected total loss 
\begin{equation}
\label{eq:multi_trig}
    \mathcal{L} = \sum_j \mathcal{L}(y_j) 
\end{equation}
Conveniently, this independence is wide-spread in HEP triggers due to the inclusivity requirements detailed in Section \ref{subsec:incl_trig}.
Independence requires that (i) the detector inputs used by each $y_i$ are maximally disjoint, and (ii) parameter sharing between different $y_i$ is minimised, except when performing common tasks such as denoising or calibration.
These two criteria are exemplified by LHC jet triggers. 
The inputs are disjoint because jets do not share constituents. 
While each jet processing chain employs the same denoising and calibration algorithms, these operations act independently on each jet and target the same type of physics object.

Joint optimisation is still possible where triggers do interdepend, but the probabilistic arguments used previously are less robust and the multitask learning problem is more challenging. 
We leave an exploration of these methods for future studies, and instead focus on the conditionally independent case in this work.

\subsection{Hardware constraints}
\label{subsec:hw_constraints}
Real deployable trigger systems must meet the constraints that arise from the host hardware system. 
Hardware constraints fall into two broad categories: constraints on the complexity of the trigger algorithms, and constraints on the input and output data structure.

Algorithmic constraints come from the limited computational capacity and fixed latency budget of the host system.
Constraints on algorithmic complexity can impact physics performance.
For example, a large neural network may be highly discriminating, but may exceed the allowed latency per inference to be deployed on-chip.
We therefore seek the task-optimal estimator that meets these constraints.
In the context of machine-learning algorithms, these constraints can be met by reducing the dimension (compression) \cite{gupta2015deeplearninglimitednumerical},  \cite{NIPS1989_6c9882bb} \cite{hinton2015distillingknowledgeneuralnetwork} and bit-precision (quantisation) \cite{gupta2015deeplearninglimitednumerical},  \cite{nagel2021whitepaperneuralnetwork} of the model parameters.
Any constraint that can be applied during the training method, i.e. quantisation-aware training (QAT), is of particular interest for an end-to-end optimisation, since it can produce models that can be considered optimal under that constraint.

Data-structure constraints arise from the limited bandwidth at which data can be transmitted, for example from detector to trigger, or between trigger components within the same system. 
Bandwidth constraints can be met by encoding the data, either via compression or quantisation. 
Lossy encoding results in information loss and can therefore adversely affect the physics performance of the system.
For example, the full ATLAS event size is on the order of $\mathcal{O}$(MB) \cite{Battaglia:2006mer}. 
If the available bandwidth from the detector to the trigger system is 1 Tbps, the event size must be reduced to $\sim 3$ kB, corresponding to a reduction by $\mathcal{O}(1000)$.
Given such an extreme reduction factor, the choice of encoding rule may strongly affect the performance of the trigger. 
We therefore seek the optimal encoding rule under a given task and bandwidth constraint.
Previous work has shown that if the encoding rule is parametric and piece-wise differentiable, then the task-optimal encoding can be found via gradient descent \cite{jung2018learningquantizedeepnetworks}\cite{shlezinger2019deeptaskbasedquantization}.
We can therefore include the data encoding in the end-to-end optimisation of the trigger.

\subsection{Choosing an objective}
The choice of loss function $\mathcal{L}$ depends on the intended physics objective.
A standard approach would be to assign $c=0$ to background events and $c=1$ for events containing rare signals, framing the optimisation task as binary event classification.
For binary event classification, discrimination power is measured by the binary cross-entropy (BCE) loss
\begin{equation}
    \mathcal{L}_{BCE}[q] = c\log q + (1-c)\log(1-q),
\end{equation}
where the smooth monotone bijection 
\begin{equation}
    q=f(y)\in[0, 1]
    \label{eqn:posterior}
\end{equation} estimates the signal posterior $p(c=1|x)$ \cite{10.1162/neco.1991.3.4.461}.
However, the ideas presented in Sections \ref{subsec:trig_opt}-\ref{subsec:hw_constraints} apply to many other training objectives.
This includes unsupervised classification techniques \cite{PhysRevD.103.092007}\cite{Bardhan_2024} such as contrastive learning \cite{Metzger_2025}, anomaly detection \cite{Govorkova_2022} and one-class classification (OOC) \cite{caron2025universalanomalydetectionlhc}, or weakly-supervised classifcation techniques \cite{Bardhan_2024} such as DROCC \cite{goyal2020droccdeeprobustoneclass} and classification with noisy labels \cite{song2022learningnoisylabelsdeep}.
Event reconstruction can be optimised using standard distance metrics such mean square error (MSE), or more sophisticated point-cloud distances such as the Chamfer or Earth-Mover's distances \cite{fan2016pointsetgenerationnetwork}.

\section{Experiments}
\label{sec:experiments}

\begin{figure}[h]
    \centering
    \includegraphics[width=\linewidth]{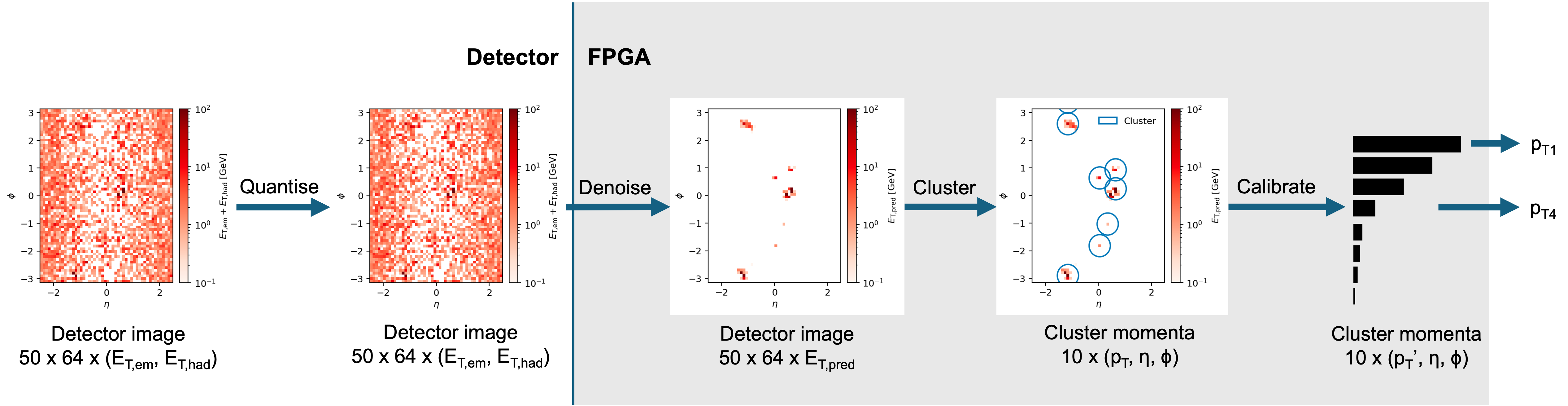}
    \caption{Diagram of jet trigger showing algorithms and intermediate data structures.}
    \label{fig:block_diagram}
\end{figure}

As an example application for the concepts introduced in this paper, we implement a realistic jet trigger based on that proposed for the ATLAS hardware trigger system at the HL-LHC.
Optimising the selection efficiency of jet triggers is especially challenging, as detector noise creates fake jets and induces clustering errors.
This trigger is based on detector image data and is made up of four algorithms.
First, the detector images are quantised and sent from the detector to an FPGA for further processing.
The images are denoised and clustered into jets.
These jets are calibrated before transverse-momentum thresholds are applied for the final trigger decision.
A block diagram of the jet trigger, showing algorithms and intermediate data structures, is shown in Fig. \ref{fig:block_diagram}.

The HL-LHC ATLAS hardware trigger is an extremely latency-constrained system.
The system will have $\mathcal{O}(10\text{ $\mu$s})$ from collision to trigger decision \cite{tdr}.
We include two mock hardware constraints, covering the two types of constraint described in Section \ref{subsec:hw_constraints}. 
The first limits the data transfer bandwidth from detector to FPGA to 2~Tbps.
The second limits the time to denoise the full image to 10 $\mu$s.
We reserve a full set of realistic hardware constraints for future work.

\begin{table}[h!]
    \caption{Hardware constraints included in this work.}
    \begin{tabular}{@{}lllr@{}}
        \toprule
        Type & Specific constraint & Applies to \\
        \midrule
        Data & $\leq2$ Tbps bandwidth from detector to FPGA  & Quantisation  \\ 
        Algorithmic & Denoise full image in $\leq10$ $\mu$s & Denoising \\
        \botrule
    \end{tabular}
    \label{tab:hw_constraints}
\end{table}
In this paper, we focus on optimising a single-jet and 4-jet trigger, which play a crucial role in the potential detection of Higgs pair production \cite{tdr}.
The triggers are constructed to respect the independence criteria set out in Section \ref{subsec:multi_trig}.
We perform two optimisations, one using a sequential framework and the other using the end-to-end framework proposed in this work.
Both optimisations start from the same initial conditions, meaning that any differences are purely the result of the optimisation strategy and associated training objective.
The frameworks are evaluated by comparing the event selection efficiencies for a range of rare physics processes of high interest, including Higgs pair production.
We also compare the physics objects output by each intermediate algorithm. 

This section is structured as follows; in Section~\ref{sec:jargon}, we give any LHC-specific definitions and assumptions, and frame this use-case in a wider non-LHC context.
Section~\ref{sec:data} details the simulated datasets used in this work.
In Section~\ref{sec:training}, we describe the sequential and end-to-end training configurations in detail.
In Sections~\ref{sec:quantisation}-\ref{sec:calibration}, we describe in detail the configurations of each trigger algorithm, and compare the algorithm-specific results from each optimisation framework.
Lastly, Section~\ref{sec:physics_perf} compares the overall physics performance of each optimisation framework.

\subsection{Assumptions}
\label{sec:jargon}

The design choices in this experiment are inspired by the ATLAS hardware trigger system planned for the HL-LHC era.
We represent momenta in the coordinate system $\vec{p}=(\pt, \eta, \phi)$, widely used at the LHC, where $\pt$ stands for the momentum transverse to the beam line, $\eta$ is pseudorapidity (linked to the polar angle $\theta$ via $\eta = -\log [\tan (\theta/2)]$), and $\phi$ is the azimuthal angle. 
Angular distances are measured using $\Delta R=\sqrt{(\Delta \eta)^2 + (\Delta \phi)^2}$.

The inputs to the trigger algorithms are detector images.
The detector in this study consists of two subdetector layers, based on the electromagnatic and hadronic calorimeters used in ATLAS respectively \cite{Boumediene:2017nhu}.
The detector samples the transverse energy $\Et$ with pixel size of $\Delta\eta \times \Delta\phi = 0.1\times\pi/32$, where $\Et=E/\cosh{\eta}$.
The two sub-detectors return a separate $\Et$ value for each pixel, which we label $\Etem$ and $\Ethad$ respectively.
The detector covers the range $|\eta|<2.5$ and $-\pi\leq\phi<\pi$, such that each image contains $50\times 64$ pixels.
Jets are defined as clusters of momenta summed within $\Delta R<0.4$.
We assume that all pixels and clusters are massless, such that $\Et=\pt$.
Jets are $\pt$-sorted in descending order, such that $\ptone$ and $\ptfour$ correspond to the jets with highest and fourth-highest $\pt$.
At the HL-LHC, each event in the ATLAS and CMS detectors will contain of order $200$ simultaneous proton-proton collisions \cite{tdr}, the overwhelming majority of which are low-energy interactions, called pile-up.
Pile-up will be the dominant source of noise in the ATLAS and CMS detectors at HL-LHC. 

\subsection{Datasets}
\label{sec:data}
\begin{figure}[h!]
    \centering
    \begin{subfigure}[]{0.49\textwidth}
        \centering
        \includegraphics[width=\textwidth]{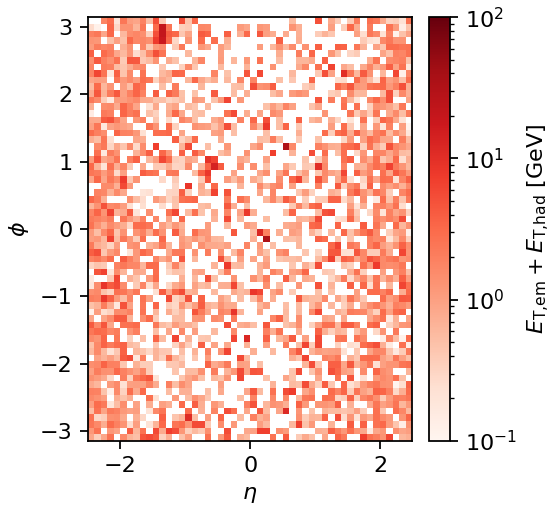}
        \caption{$\pu=200$}
    \end{subfigure}
    \hfill
    \begin{subfigure}[]{0.49\textwidth}
        \centering
        \includegraphics[width=\textwidth]{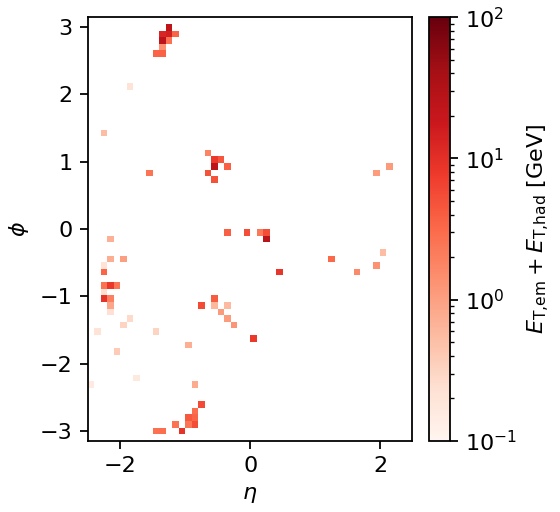}
        \caption{$\mu=0$}
    \end{subfigure}
    \caption{Example \ttbar\ detector image at $\pu=200$ (left) and $\mu=0$ (right).}
    \label{fig:event_displays}
\end{figure}

Samples of simulated proton-proton collision events are generated at $\sqrt{s}=14$ TeV using the open-source \textsc{HEPData4ML} framework.
We use the \textsc{Pythia8.3} package \cite{Bierlich:2022pfr} for generating the initial proton-proton interaction and subsequent decays of unstable particles.
The \textsc{Delphes} framework \cite{deFavereau:2013fsa} is used to simulate the ATLAS detector, customised to reproduce the experimental design in Section \ref{sec:jargon}.
A background-only sample of 500k dijet QCD interactions is generated.
Samples of four rare signal processes are also generated, each containing 100k events.
These rare processes are combined with equal mixture weights into a single inclusive signal sample for training, covering a variety of jet-only event topologies.
An additional auxiliary sample of 1M soft QCD interactions is generated and used to model pile-up effects.
Pile-up effects are modelled by randomly sampling and overlaying $\mu$ soft QCD interactions per event.
The detector simulation is run twice for each signal and background event, once with $\mu \sim \text{Pois(200)}$ to model HL-LHC conditions, and once with $\mu=0$ to provide truth labels in the absence of noise.
Two example detector images are shown in Fig. \ref{fig:event_displays}.

The momenta of generator-level stable particles in $\mu=0$ events are used to reconstruct so-called truth jets, using the anti-$k_t$ jet clustering algorithm \cite{Soyez:2008pq} with $R=0.4$. Truth jets with $\pt$ $>20$~GeV are used as ground-truth for the jet calibration.
The detector images at $\mu=0$ are used as ground-truth for the denoising task.
A summary of the samples used and their different contributions into the training process can be found in Table \ref{tab:Samples}.
Each sample undergoes a randomised 50\%-50\% train-test split, with 20\% of the training events set aside for validation.

We stress that the success of our end-to-end framework is not dependent on the ATLAS-specific design choices, datasets and object definitions.
In order to emphasise the wider applicability of our framework, we henceforth switch to a set of experiment-agnostic terms.
These are summarised in Table~\ref{tab:glossary} for ease of reference.
Jets will be referred to as clusters, and the two calorimeter sub-detectors will be collectively referred to as the detector.
Pile-up effects will be referred to simply as noise.
We refer to the fraction of signal and background events accepted by the trigger as the true-positive rate (TPR) and false-positive rate (FPR) respectively.

\begin{table}[h!]
    \caption{Glossary of experiment-specific terms.}%
    \begin{tabular}{@{}lllr@{}}
        \toprule
        Term & Definition \\
        \midrule
        Detector & ATLAS electromagnetic and hadronic calorimeters \\
        Detector image & $50\times 64$ image of $\Et$ \\
        Pixel & $(\Etem, \Ethad)$ for a single unit of size $\Delta\eta \times \Delta\phi = 0.1\times\pi/32$ \\
        Quantisation & On-detector discretisation of $\Et$ values\\
        Noise & $\sim200$ soft QCD collisions overlaid on each image\\
        Cluster & Small-radius ($R=0.4$) jet\\
        Calibration & Calibrate calorimeter jet $\pt$ using overlapping truth jet $\pt$  \\
        Triggers & Highest ($\ptone$) and fourth-highest ($\ptfour$) jet $\pt$ \\
        \botrule
    \end{tabular}
    \label{tab:glossary}
\end{table}

\begin{table}[h]
    \caption{Generated samples for the use case example. VBF and ggF refer to the vector-boson and gluon-gluon fusion production modes. 
    Higgs + missing energy refers to Higgs boson production in assocation with a $Z$-boson which decays to neutrinos.
    Top quark pair production refers to the simultaneous production of a top and anti-top pair. Higgs bosons decay into $b$-quarks, and top quarks decay hadronically. Finally, we define di-jet QCD as the inclusive production of two jets.}\label{dataConfTable}%
    \begin{tabular}{@{}lllll@{}}
        \toprule
        Name & Process & Pile-up $\langle\mu\rangle$ & Category & Events\\
        \midrule
        VBF Higgs pair-production & \vbfhh & 0/200  & Signal  & 100,000  \\
        ggF Higgs pair-production & \ggfhh & 0/200 & Signal  & 100,000  \\
        Higgs + missing energy & \hzbbvv  & 0/200 & Signal & 100,000  \\
        Top quark pair-production & \ttbar & 0/200 & Signal & 100,000 \\
        Di-jet QCD & $jj,\, j\in\{g, q\}$ & 0/200 & Background & 500,000 \\
        \botrule
    \end{tabular}
    \label{tab:Samples}
\end{table}

\subsection{Training}
\label{sec:training}
We enable optimisation via gradient descent by implementing the trigger as a single model within the \textsc{TensorFlow} package.
Two models are trained, using the sequential and end-to-end optimisation frameworks.
Both optimisations use the same initial conditions, reusing the same random splitting of the dataset, parameter initialisation and hyperparameters in both trainings.
The training loop configuration is also kept the same, and is detailed in Table \ref{tab:TrainConfig}.

\begin{figure}[h!]
    \centering
    \includegraphics[width=0.5\linewidth]{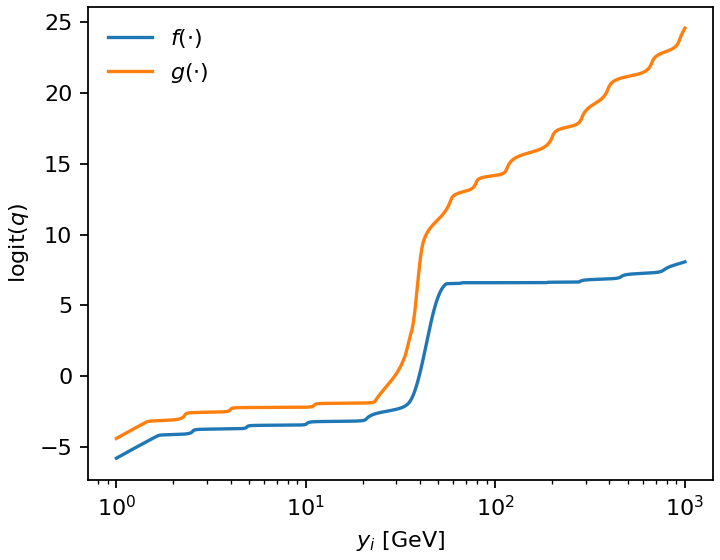}
    \caption{Visualisation of the trained bijectors $f(\cdot)$ and $g(\cdot)$ in the end-to-end model.}
    \label{fig:splines}
\end{figure}

In the sequential framework, each algorithm is optimised in order, from quantisation first to calibration last.
The parameters of any upstream algorithms are frozen during each training.
Each algorithm is trained to minimise its corresponding loss function, which can be found in Table~\ref{tab:SequentialTraining}.
\begin{table}[h!]
    \caption{Training loss function configurations.}
    \begin{tabular}{@{}lllr@{}}
        \toprule
        \multirow{2}{*}{Algorithm} & \multicolumn{2}{c}{Loss function} \\
        & Sequential & End-to-end\\
        \midrule
        Quantisation & MSE per-pixel  & \multirow{3}{*}{$\mathcal{L}_\text{e2e}$ (Eq. \ref{eq:End2EndLoss})}\\ 
        Denoising & MSE per-pixel  \\ 
        Calibration & $D_C$ (Eq. \ref{eq:CalibrationLoss}) \\
        \botrule
    \end{tabular}
    \label{tab:SequentialTraining}
\end{table}
In our end-to-end framework, all the trainable algorithms are optimised in a single training.
We do this by summarising physics performance as a single loss function, and minimising this loss via gradient decent on all algorithm parameters simultaneously, in line with the theory discussed in Section \ref{sec:theory}. 
The loss function is defined by combining Eqns. \ref{eq:bce_w_calib} and \ref{eq:multi_trig}, containing two disjoint BCE terms and a convex calibration term $D_C$:
\begin{equation} \label{eq:End2EndLoss}
    \mathcal{L}_\text{e2e} = 
    \mathcal{L}_{BCE}[{f(\ptone})]+ 
    \mathcal{L}_{BCE}[{g(\ptfour)}] + \alpha  D_C 
    + \text{penalty terms},
\end{equation}
where $\alpha$ is a user tunable parameter. 
In setting $\alpha=5\times 10^{-4}$, we explicity optimise for event classification.
While $\alpha>0$ ensures that clusters are calibrated, $\alpha\ll0.5$ ensures the conditional variance in the calibration is optimised for classification rather than perfect reconstruction. 
The bijections $f(\cdot)$ and $g(\cdot)$ are implemented as rational-quadratic splines \cite{durkan2019neuralsplineflows}, each with 16 knots. 
The splines are smoothed by penalising the second derivative $\lambda f^{''}(\cdot)$ where $\lambda=10^{-3}$, thus moderating the gradients propagated to the upstream algorithms.
The trained bijectors are shown in Fig. \ref{fig:splines}, showing how the splines map $\ptone$ and $\ptfour$ to their respective posteriors as per Eqn. \ref{eqn:posterior}.
Note that the splines are only needed during optimisation and are not deployed in the trigger system.

\begin{table}[h]
    \caption{Model training loop configuration}
    \begin{tabular}{@{}lllr@{}}
        \toprule
        Parameter & Value \\
        \midrule
        Optimizer & Adam \\
        Learning Rate (LR) & $10^{-3}$, halved every 250 epochs\\
        Epochs & 1000 \\
        Batch size & 128 \\
        Checkpointing & After lowest validation loss \\
        \botrule
    \end{tabular}
    \label{tab:TrainConfig}
\end{table}

\subsection{Quantisation}
\label{sec:quantisation}
The detector images are discretised by a nearest-value quantiser $Q$ such that
\begin{equation}
    Q(x) = c_j, \, j=\arg\min_i |x-c_i|, 
\end{equation}
where $c$ is a vector of possible output values sorted in ascending order.
Each quantiser can be parametrised by a lowest-value $c_0$ and a vector of bin widths ${\Delta_i}=c_{i+1}-c_i$.
The widths $\Delta_i$ are treated as independent learnable parameters and are allowed to vary during training.
During training, we fit a differentiable smooth approximation $\bar Q$ such that
\begin{equation}
    \bar Q(x) = \sum_i c_i \,\text{softmax}(-T |x-c_i|),
\end{equation}
where $T\gg1$ controls the degree of smoothing.
This approach is based on the task-aware quantisers proposed in \cite{shlezinger2019deeptaskbasedquantization}.
Note that $\bar Q \rightarrow Q$ as $T\rightarrow \infty$; here we set $T=50$.
We revert to a discrete quantiser $Q$ using the same $c$ during inference.
\begin{figure}
    \centering
    \includegraphics[width=0.5\linewidth]{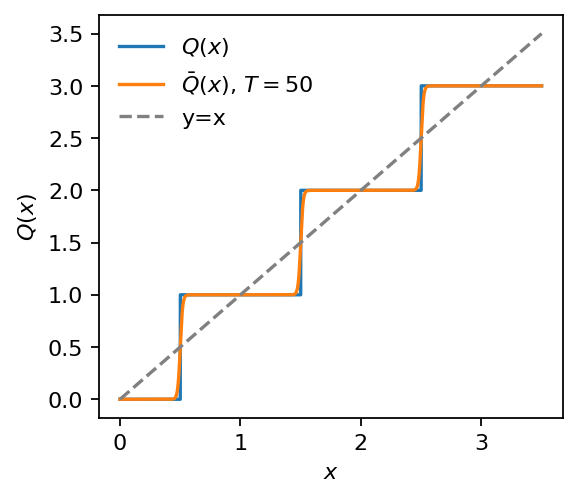}
    \caption{Example 2-bit nearest-value quantiser with lower limit $c_0=0$ and bin widths $\Delta_i=1$.}
    \label{fig:example_quantiser}
\end{figure}

Here $\Etem$ and $\Ethad$ are quantised independently by $Q_\text{em}$ and  $Q_\text{had}$ respectively.
We set the number of possible output values to $256$, such that $Q(x)$ can be encoded using 1 byte.
This gives an event size of 6.4 kB, and thus a corresponding data rate of 2 Tbps for a 40 MHz event rate.
Since $\Et\geq0$, we fix the lower limit $c_0=0$ GeV.
The widths $\Delta_i$ are initialised at 0.4 GeV for both $Q_\text{em}(x)$ and $Q_\text{had}(x)$, giving an initial upper limit of $c_{255}=102$ GeV in both cases.
Since we expect $\Delta_i$ to vary smoothly with $\Et$, we apply a regularisation term to the loss, defined
\begin{equation}
    \mathcal{L}_{\Delta} = \lambda\frac{\sum_i|\Delta_{i+1}-\Delta_i|}{\sum_i\Delta_{i}},
\end{equation}
with $\lambda=10^{-3}$.

In the sequential optimisation, each of $Q_\text{em}(x)$ and $Q_\text{had}(x)$ minimise the widely used MSE between input $\Et$ and the quantised output \cite{shlezinger2019deeptaskbasedquantization}\cite{Guglielmo_2021}.
In the end-to-end optimisation, $Q_\text{em}(x)$ and $Q_\text{had}(x)$ are allowed to evolve simultaneously with the other algorithms in order to minimise the joint loss function in Eqn. \ref{eq:End2EndLoss}.

\begin{figure}[h!]
    \centering
    \begin{subfigure}[b]{0.49\textwidth}
        \centering
        \includegraphics[width=\linewidth]{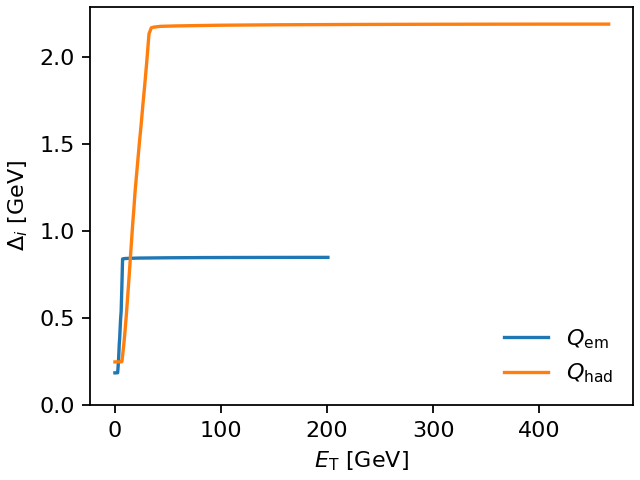}
        \caption{Sequential}
        \label{}
    \end{subfigure}
    \begin{subfigure}[b]{0.49\textwidth}
        \centering
        \includegraphics[width=\linewidth]{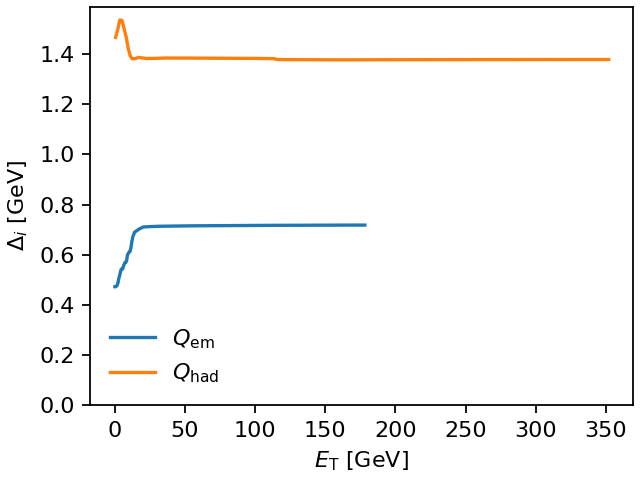}
        \caption{End-to-end}
        \label{}
    \end{subfigure}
    \caption{Bin widths $\Delta_i$ as a function of input $\Et$.}
    \label{fig:quantisers}
\end{figure}

The trained quantisers are shown in Fig. \ref{fig:quantisers}.
The abundance of low-$\Et$ pixels compresses $\Delta_i$ at low-$\Et$ for the sequentially optimised quantiser.
The end-to-end quantiser favours more uniform $\Delta_i$ over a narrower $\Et$ range.
The MSE-optimal sequential quantiser must provide good $\Et$-resolution across the full input $\Et$ range, whereas the end-to-end quantiser can prioritise $\Et$-resolution in the $\Et$ range which are more likely to inform signal-background discrimination and calibration. 
\subsection{Denoising}
The denoising algorithm uses a convolutional neural network (CNN) based on the algorithm proposed in \cite{ATL-UPGRADE-PUB-2025-002}. 
The CNN predicts a pair of segmentation weights $w=(w_\text{em}, w_\text{had})\in [0, 1]$ for each pixel.
The denoised image is given by
\begin{equation}
    \Etpred = w_\text{em}\, \Etem + \,w_\text{had}\, \Ethad
\end{equation}
Predicting $w=0$ signifies that the pixel contains mostly noise, while $w=1$ signifies that the pixel is likely to form a real cluster.
The image is cyclic-padded at the $\phi$-edge and zero-padded at the $\eta$ edge by 1 pixel.
The first layer of the CNN is a 2D $(\eta, \phi)$-symmetric depthwise convolutional layer with a kernel size of $3\times3$, depth multiplication factor of 4 and ReLU activation.
The resulting per-pixel vector of convolved features is concatenated with the $\log{|\eta|}$ coordinate of the pixel.
This vector is then passed through 2 dense layers, each with 32 neurons and the ReLU activation.
The predicted $w$ is extracted by a final 2-neuron dense layer with the hard-sigmoid activation.
The CNN is equivariant under translation in $\phi$ and under parity inversion $(\phi\rightarrow\pi-\phi, \,\eta\rightarrow-\eta)$.

In the sequential training, the CNN is trained to minimise the MSE per pixel, using $\Etem + \Ethad$ in the noise-free ($\mu=0$) image as a target.
In the end-to-end case, the CNN is allowed to evolve simultaneously with the other algorithms in order to minimise the joint loss function in Eqn. \ref{eq:End2EndLoss}.

The CNN is implemented using QAT via the \textsc{QKeras} package \cite{Coelho_2021}.
The precision of the CNN parameters are uniformly set to 10~bits.
The CNN is compiled into FPGA firmware using the \textsc{HLS4ML} package \cite{Coelho_2021}.
The CNN firmware is able to process a new pixel every 3 ns, which scales to 9.8 $\mu$s for the full $50\times64$-pixel image, assuming no parallel processing or time-multiplexing.

\begin{figure}
    \centering
    \begin{subfigure}[]{\textwidth}
        \centering
        \includegraphics[width=\textwidth]{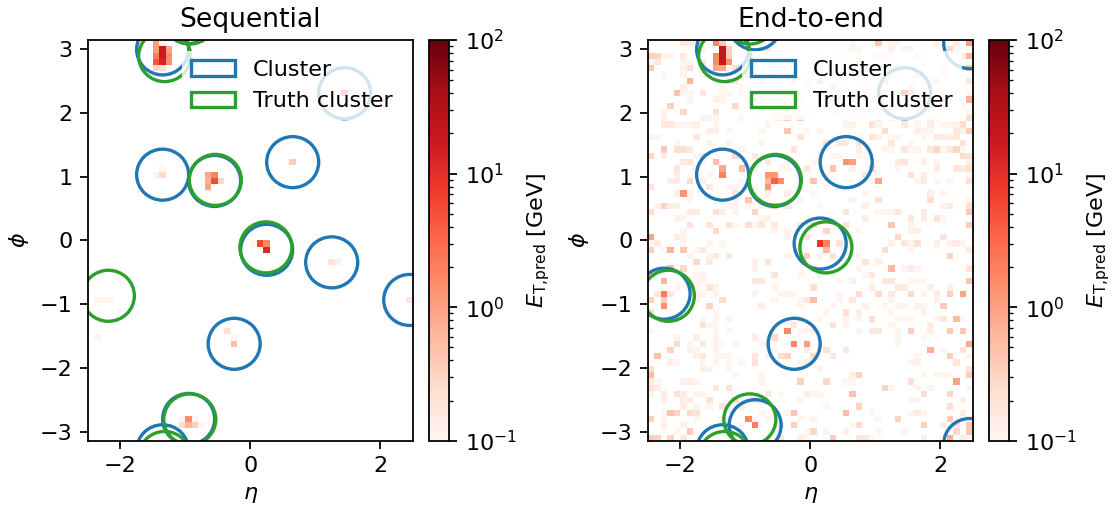}    
    \end{subfigure}
    \caption{Example denoised \ttbar\ event for the sequential (left) and end-to-end (right) model, showing truth and reconstructed clusters.}
    \label{fig:suppression}
\end{figure}

Fig. \ref{fig:suppression} shows an event display after the denoising stage of each model.
The events shown are the same as in Fig \ref{fig:event_displays}.
Both models suppress noise and improve the identifiability of truth clusters.
While the end-to-end model admits more noise pixels, these pixels are more strongly $\Et$-suppressed, achieving a greater $\pt$ separation between background and signal clusters.
This highlights the importance of end-to-end task-optimality; there is no guarantee that minimising any per-pixel loss function will produce optimal event-level classification or reconstruction.

\subsection{Clustering}
The denoised calorimeter images are clustered using a cone algorithm. 
The algorithm is inspired by the jet clustering algorithm currently in use by the ATLAS hardware trigger \cite{Villhauer:2719536}.
First, we locate seeds, which are local $\Etpred$ maxima within a rounded-off square window of $P\times P$ pixels.
The cluster $\pt$ is then taken as the sum of $\Etpred$ within the window.
By setting $P=9$, we approximate the anti-$k_t$ algorithm with $\Delta R=0.4$ used to build truth clusters.
The cluster $\eta$ and $\phi$ are taken as the coordinates of the seed.
The algorithm returns a $\pt$-sorted list of momentum vectors for the 10 highest-$\pt$ clusters in each event. 
The list is zero-padded for events with less than 10 clusters.
The cone algorithm has no trainable parameters and does not vary during training.

\subsection{Calibration}
\label{sec:calibration}
The $\pt$ of the clusters is calibrated to account for noise and the $\Et$-response of the detector. 
Each cluster is calibrated independently.
The calibration $H$ transforms the cluster $\pt$, conditioned on $|\eta|$ such that
\begin{equation}
    \pt \rightarrow \pt^\prime = H(\pt,|\eta|),
\end{equation}
where $H$ is $\pt$-monotonic.
Cluster $\eta$ and $\phi$ are unchanged.
The $\pt$-monotonicity of $H$ is a core design requirement of LHC triggers, since preserving $\pt$-ordering avoids instabilities in threshold-based selections.
However, it places a strong functional constraint on $y$; the optimal model may therefore display some conditional bias in the calibration, as discussed in Section \ref{subsec:calibration}.
Monotonicity assumes that the minimiser of $\E[D_C | \pt]$ is a monotonic function of $\pt$; however, this may not be true in general. 

In this work, we implement $H$ as a dense neural network, using the partially-monotonic dense layers proposed by \cite{runje2023constrainedmonotonicneuralnetworks}. 
The network is constrained to be monotonic in $\pt$ but not $|\eta|$.
The inputs $(\pt,|\eta|)$ are log-normalised before being passed to the network. 
The neural network has two hidden layers of 64 neurons with the softplus activation and an output layer of 2 nodes.
The first output node is passed through an exponential activation, mapping the output to $\pt$ space.
The second is passed through a sigmoid activation, acting as a gate on $[0, 1]$.
The final $\pt$ prediction is the product of the exponentiated and the gated outputs.
Multiplying by the gated output allows the calibration to explicitly suppress the low-$\pt$ clusters which may be associated with background or noise.
The result is a calibration curve which is $\pt$-monotonic, non-negative and can drop sharply to zero below a certain input $\pt$.
We implement the following calibration loss
\begin{equation} \label{eq:CalibrationLoss}
    D_C =  \frac{\sum_{t, r} M_{tr} \, d_{tr}^2}{\sum_{t, r} M_{tr}},
\end{equation}
where the matrix $d_{tr} = p_{\text{T},t}-p'_{\text{T},r}$ and the matrix ${M_{tr}}$ is defined as
\begin{gather}
    M_{tr}^{(0)} = \textbf{1}\left(\Delta R_{tr} < R_\text{max}\right), \\
    {M_{tr}} = M_{tr}^{(0)} \, B_r \, B_t,
\label{eqn:matching}
\end{gather}
where $B_{t} =\textbf{1}\left(\sum_{r'}M_{tr'}^{(0)}=1\right)$ and the indices $t$ and $r$ run over the truth and reconstructed clusters respectively. 
Here $\textbf{1}(\cdot)$ is the indicator function.
The $D_C$ loss minimises the MSE on $\pt$ between truth and reconstructed clusters that coincide within $\Delta R<R_\text{max}$, with the vectors $B$ enforcing unique pairing. Here we set $R_\text{max}=0.3$ in line with \cite{ATL-UPGRADE-PUB-2025-002}. 

\begin{figure}
    \centering
    \begin{subfigure}[]{0.49\textwidth}
        \centering
            \includegraphics[width=\linewidth]{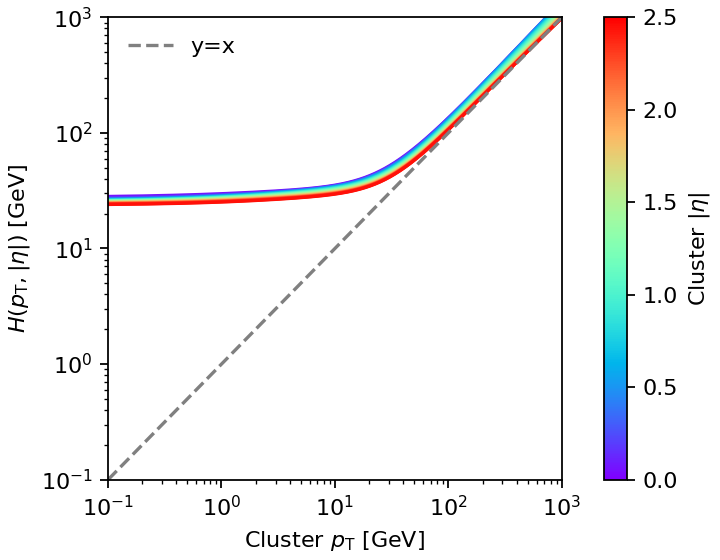}
        \caption{Sequential}
    \end{subfigure}
    \begin{subfigure}[]{0.49\textwidth}
        \includegraphics[width=\linewidth]{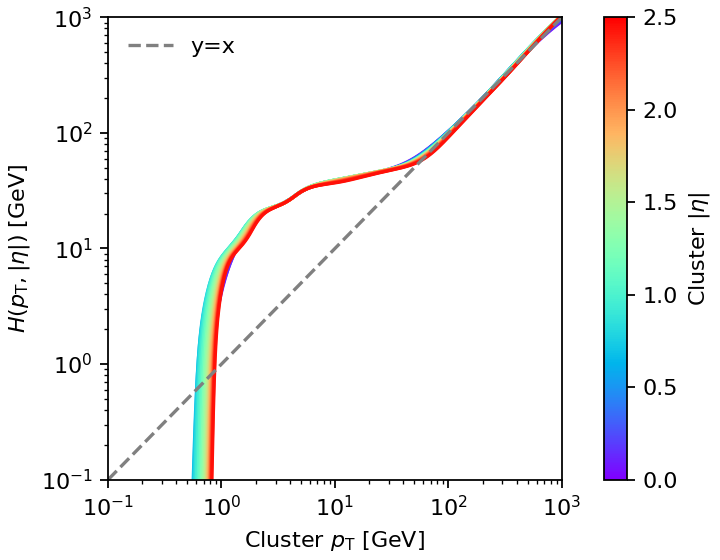}
        \caption{End-to-end}
    \end{subfigure}
    \caption{Response of the calibration neural network $H$ in the sequential (a) and end-to-end (b) models.}
    \label{fig:jes_response}
\end{figure}

\begin{figure}[h]
    \centering
    \includegraphics[width=0.7\linewidth]{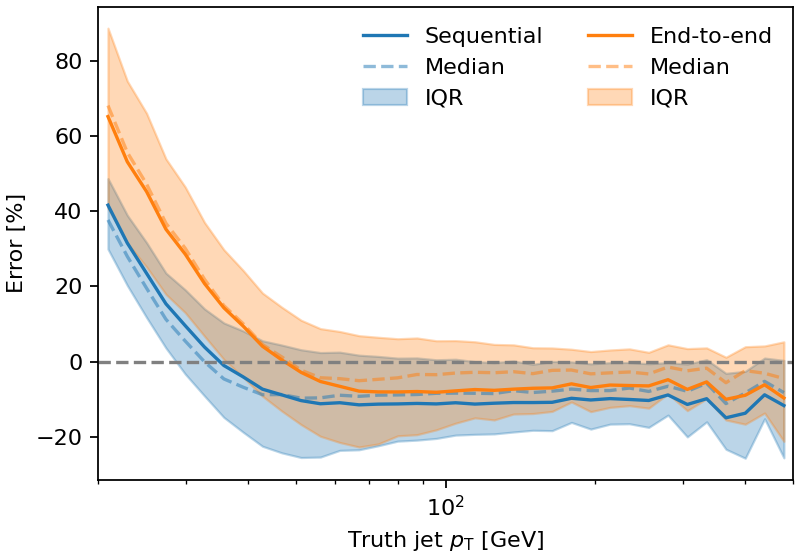}
    \caption{Percentage error on cluster $\pt$ in the test \ttbar\ sample as a function of truth cluster $\pt$.}
    \label{fig:pt_residual}
\end{figure}

Fig. \ref{fig:jes_response} shows the response of $H$ as a function of cluster $\pt$.
In the end-to-end case, cluster $\pt$ response turns on sharply between $0.5-2$ GeV, depending on $|\eta|$.
The calibration has learnt an optimal $|\eta|$-dependent $\pt$ threshold for cluster seeding, something which is not possible in the sequential framework.
To evaluate the calibration, truth and reconstructed clusters are matched uniquely with the criteria in Eqn. \ref{eqn:matching}, and the percentage error between reconstructed and truth $\pt$ is taken.
This error is shown in Fig. \ref{fig:pt_residual} as a function of truth $\pt$ for the test \ttbar\ sample, with the median and interquartile range (IQR) also shown.
Both models show conditional biases at low-$\pt$ due to the $\pt$-monotonicity constraint on $H$.
The calibration of the end-to-end model has a higher bias and variance than the sequential model below a truth $\pt$ of 100 GeV, and has lower bias and variance above 100 GeV.
This highlights a key advantage of the end-to-end approach, which has the freedom to balance noise suppression and $\pt$-precision in order to maximise discrimination. 
The end-to-end model has learnt that suppressing the $\pt$ of certain background clusters is worth sacrificing some $\pt$-resolution for, an advantage that the sequential model cannot leverage. 

\subsection{Physics performance}
\label{sec:physics_perf}
\subsubsection{Main triggers}
\begin{figure}[h!]
    \centering
    \begin{subfigure}[]{0.49\textwidth}
        \centering
         \includegraphics[width=\textwidth]{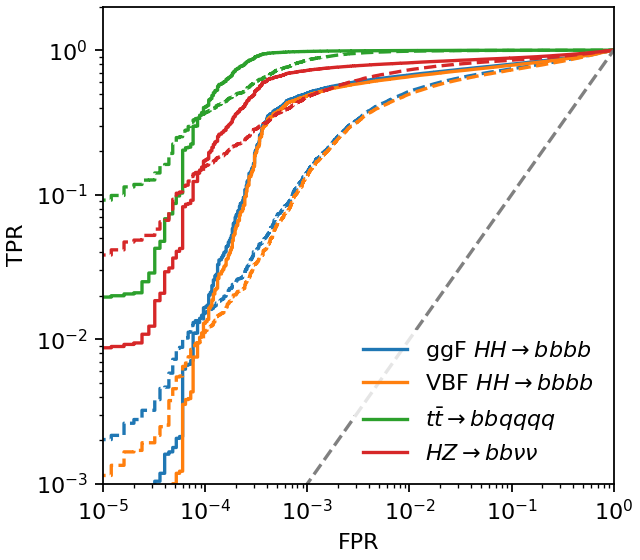}
         \caption{$\ptone$}
    \end{subfigure}
    \begin{subfigure}[]{0.49\textwidth}
        \centering
        \includegraphics[width=\textwidth]{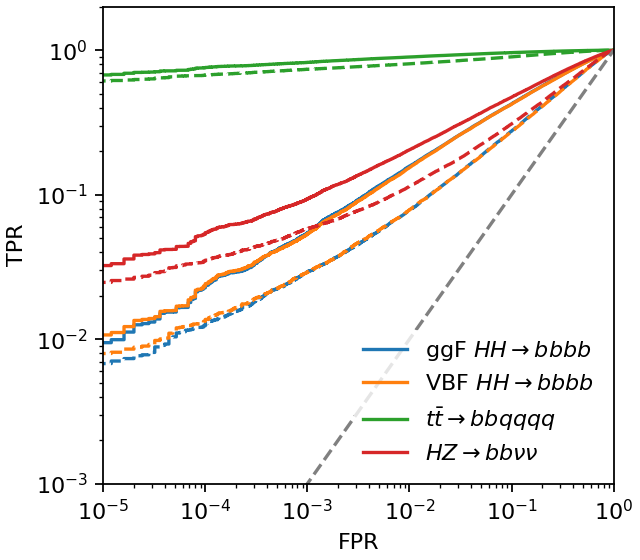}
         \caption{$\ptfour$}
    \end{subfigure}
    \begin{subfigure}[]{0.49\textwidth}
        \centering
         \includegraphics[width=\textwidth]{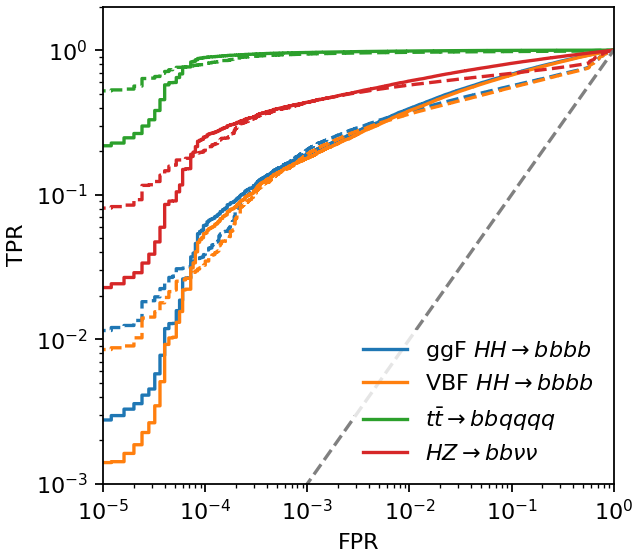}
         \caption{\Ht}
    \end{subfigure}
    \begin{subfigure}[]{0.49\textwidth}
        \centering
         \includegraphics[width=\textwidth]{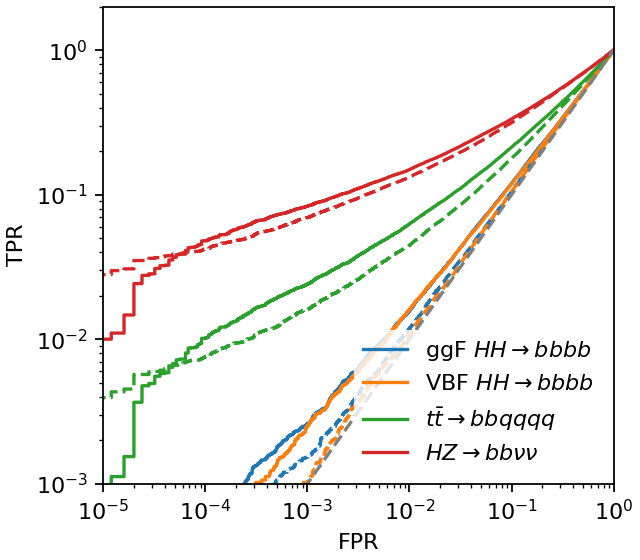}
         \caption{\MHt}
    \end{subfigure}
    \caption{ROC curves for each signal and trigger observable for the sequential (dashed lines) and end-to-end models (solid lines).}
    \label{fig:roc_curves}
\end{figure}

We quantify the signal-background discrimination of each model using the area under the ROC curve (AUC) and the TPR at FPR $10^{-3}$ for each signal.
The values are tabulated in Table \ref{tab:ptj1_ptj4_performance}.
The complete set of ROC curves are shown in Fig. \ref{fig:roc_curves}.
The end-to-end model greatly improves the selection efficiency over $\ptone$ and $\ptfour$ for all signals.
Differences in performance between signals stem from differences in the marginal distributions over $\ptone$ and $\ptfour$. 

\begin{table}[h!]
   \caption{Performance assessment of the sequential and end-to-end models for each signal for $\ptone$ and $\ptfour$.}
   \centering
   \begin{tabular}{@{}llcccc@{}}
       \toprule
       Signal & Model & \multicolumn{2}{c}{AUC} & \multicolumn{2}{c}{TPR @ FPR $10^{-3}$} \\
       & & $\ptone$ & $\ptfour$ & $\ptone$ & $\ptfour$ \\
       \midrule
       \multirow{2}{*}{\ggfhh}
       & Sequential & 0.88 & 0.66 & 0.14 & 0.029 \\
       & End-to-end & \textbf{0.90} & \textbf{0.75} & \textbf{0.50} & \textbf{0.054} \\
       \midrule
       \multirow{2}{*}{\vbfhh}
       & Sequential & 0.87 & 0.66 & 0.13 & 0.029 \\
       & End-to-end & \textbf{0.89} & \textbf{0.75} & \textbf{0.48} & \textbf{0.053} \\
       \midrule
       \multirow{2}{*}{\ttbar}
       & Sequential & 1.00 & 0.96 & 0.85 & 0.74 \\
       & End-to-end & \textbf{1.00} & \textbf{0.99} & \textbf{0.98} & \textbf{0.82} \\
       \midrule
       \multirow{2}{*}{\hzbbvv} 
       & Sequential & 0.93 & 0.68 & 0.47 & 0.058 \\
       & End-to-end & \textbf{0.94} & \textbf{0.77} & \textbf{0.72} & \textbf{0.093} \\
       \botrule
   \end{tabular}
   \label{tab:ptj1_ptj4_performance}
\end{table}

\subsubsection{Other triggers}
We also sum the $\pt$ of the calibrated clusters for each event, taking both the scalar sum (\Ht) and vector sum (\MHt).
While not explicitly included in the optimisation due to conditional-independence constraints, these multi-object triggers may benefit from any improvement in noise suppression and calibration.
The AUC and TPR at FPR $10^{-3}$ are tabulated in Table \ref{tab:ht_htmiss_performance}.
The end-to-end model generally improves signal-background discrimination, despite not being directly optimised on \Ht\ and \MHt.

\begin{table}[h!]
   \caption{Performance assessment of the sequential and end-to-end models for each signal for \Ht\ and \MHt.}
   \centering
   \begin{tabular}{@{}llcccc@{}}
       \toprule
        Signal & Model & \multicolumn{2}{c}{AUC} & \multicolumn{2}{c}{TPR @ FPR $10^{-3}$} \\
        & & \Ht & \MHt & \Ht & \MHt \\
       \midrule
       \multirow{2}{*}{\ggfhh}
       & Sequential & 0.76 & 0.51 & \textbf{0.21} & 0.0015 \\
       & End-to-end & \textbf{0.87} & \textbf{0.52} & 0.19 & \textbf{0.0026} \\
       \midrule
       \multirow{2}{*}{\vbfhh}
       & Sequential & 0.75 & 0.51 & \textbf{0.19} & 0.0010 \\
       & End-to-end & \textbf{0.86} & \textbf{0.52} & 0.18 & \textbf{0.0025} \\
       \midrule
       \multirow{2}{*}{\ttbar}
       & Sequential & 0.99 & 0.57 & 0.94 & 0.016 \\
       & End-to-end & \textbf{1.00} & \textbf{0.58} & \textbf{0.96} & \textbf{0.024} \\
       \midrule
       \multirow{2}{*}{\hzbbvv}
       & Sequential & 0.82 & 0.66 & 0.43 & 0.069 \\
       & End-to-end & \textbf{0.92} & 0.66 & \textbf{0.44} & \textbf{0.083} \\
       \botrule
   \end{tabular}
   \label{tab:ht_htmiss_performance}
\end{table}

\section{Discussion}
\label{sec:discuss}
Although we use a HL-LHC jet trigger as a demonstrative example, we reiterate that the success of our end-to-end framework does not depend on the LHC- and jet-specific design choices, datasets and physics object definitions.
Our approach can be applied wherever systems are piece-wise differentiable and a global physics objective can be defined, opening up applications across the HEP domain and beyond.
We note that the LHC-oriented inclusivity and conditional-independence requirements adopted here may not apply to other domains, and can be relaxed as required by the target application.

Within an LHC context, the end-to-end framework could be used to target physics objects such as electrons, photons and taus, and be applied within an HLT context.
The ideas presented here can be applied outside the LHC to many other real-time event selection systems.
In particular, we foresee applications to trigger and data-acquisition systems in neutrino experiments \cite{Kalra_2022}\cite{dune_daq} and imaging atmospheric Cherenkov telescopes (IACTs) \cite{caroff2023realtimeanalysisframework}.

We also stress that the end-to-end framework is compatible with many machine-learning tasks, not just supervised classification.
In particular, we foresee the use of weakly supervised techniques, especially in applications where high quality simulations are not available.

While the hardware constraints imposed in this paper are illustrative, they provide a proof-of-concept that limitations on both data bandwith and model compression can be incorporated into an end-to-end optimisation.
In particular, we highlight that the trainable quantiser optimised in this work is the first example of a HEP detector readout being optimised under a  global physics objective.
Our end-to-end framework could be used to link and jointly optimise many previously disconnected systems and tasks, opening up many potential applications.
For example, the autoencoder used to readout the CMS high-granularity calorimeter \cite{Guglielmo_2021} could be optimised for global event reconstruction, instead of minimising the local reconstruction error.

Lastly, we remphasise that sequentially optimising for algorithm-specific local objectives can lead to highly sub-optimal overall physics performance.
We observe a remarkable performance gap between the sequential and end-to-end strategies, despite the objectives used to optimise the sequential model being widely used and physically well-motivated.
In our jet trigger example, the end-to-end optimisation increases Higgs pair selection efficiency by $2-4$, effectively extending the data-taking period of the HL-LHC by up to 40 years.
This exposes the previously hidden physics cost of the sequential approach, and demonstrates the potential benefit of end-to-end trigger optimisation for the detection of important physics.

\section{Conclusion}
In this work, we reformulated trigger design in high-energy physics as a constrained end-to-end optimisation problem. Rather than sequentially optimising modular components under local objectives, we treated the entire trigger chain, including quantisation, denoising, clustering, calibration, and selection, as a single differentiable system optimised with respect to a global physics objective.

Using a realistic HL-LHC-inspired multi-jet trigger as a case study, we demonstrated that end-to-end optimisation yields substantial gains in physics performance relative to the traditional sequential paradigm. At fixed false-positive rate, the true-positive rate for Higgs boson pair production increases by a factor $2-4$. Importantly, these gains are achieved while maintaining physically interpretable intermediate representations, enforcing monotonic calibration constraints, and satisfying illustrative hardware limitations on bandwidth and inference latency.

Our results show that, in general, locally optimal components do not combine into a globally optimal trigger. Sequential optimisation therefore incurs a previously unquantified physics cost. By contrast, the end-to-end framework allows the system to learn task-aware trade-offs, for example, between noise suppression and calibration precision, that improve overall discrimination even when individual intermediate metrics degrade.

Beyond the specific jet-trigger application presented here, the framework is broadly applicable wherever differentiable processing chains and well-defined physics objectives exist. The ability to incorporate model compression, quantisation, and data encoding directly into the optimisation opens a new avenue for co-designing physics algorithms and hardware constraints. This paradigm may be extended to other trigger objects at the LHC, to high-level trigger systems, and to real-time data acquisition challenges in neutrino experiments, astroparticle observatories, and other large-scale scientific instruments.

End-to-end optimisation thus provides both a conceptual shift and a practical toolkit for next-generation real-time event selection systems.

\section*{Acknowledgements}

We gratefully acknowledge the support of the UK's Science and Technology Facilities Council (STFC). NCH is supported by the STFC UCL Centre for Doctoral Training in Data Intensive Science  (ST/W00674X/1) and by UCL and industry funds.
DWM is supported by the National Science Foundation under Grant No. PHY-2310094.
This work has been partially funded by the Eric \& Wendy Schmidt Fund for Strategic Innovation through the CERN Next Generation Triggers project under grant agreement number SIF-2023-004.

\bibliography{sn-bibliography}

\end{document}